# Molecular techniques to characterize the *invA* genes of *Salmonella* spp. for pathogen inactivation study in composting.

SUNAR, N. M.*+, STENTIFORD, E.I.*, STEWART, D.I.* AND FLETCHER, L.A*
+ *Department of Environmental Engineering, Faculty of Engineering, University of Tun Hussien Onn Malaysia, Batu Pahat, Johor, Malaysia.*
\**Pathogen Control Engineering (PaCE) Institute, School of Civil Engineering, University of Leeds, Leeds, LS2 9JT, United Kingdom.*
+*Corresponding Author. Tel: +44 (0) 113 3432319. Email:shuhaila@uthm.edu.my*

## EXECUTIVE SUMMARY

The relatively low concentration of pathogen indicators, such *Salmonella,* in composting sometimes causes a problem with detection when using the conventional techniques. The presence of viable but non-culturable (VBNC) organisms is also a potential problem with *Salmonella* detection when using conventional techniques. In this study, the molecular approach for organism recognition, known as Polymerase Chain Reaction (PCR), was used for characterisation the *Salmonella* spp. used as inoculums in composting. This study also provides a better understand about selecting the suitable primer for *Salmonella* spp. The specificity of the probes and primers at the species level were verified by performing NCBI-BLAST2 (Basic Local Alignment Search Tool). Primers that target the *invA* gene for *Salmonella* were selected which produce fragment lengths around 285bp (shown in Figure 1). The *Salmonella* spp. solutions were tested and proved contained the sequences of *invA* gene by using several steps of molecular techniques before used it as inoculums in composting. The laboratory scale batch composting reactors were used to examine the inactivation of *Salmonella* spp. The inoculate solution of *Salmonella* was prepared by culturing *Salmonella enteritidis* (ATCC13076) in tryptone broth solution for 24 hours before adding it directly into the compost material. The reduction rate of *Salmonella* spp. was enumerated by conventional method accordingly to the standard compost guidelines (Figure 2). The laboratory scale study showed that after composting for 8 days the numbers of *Salmonella* spp. were below the limits in the UK compost standard (PAS 100) which requires the compost to be free of *Salmonella* spp.

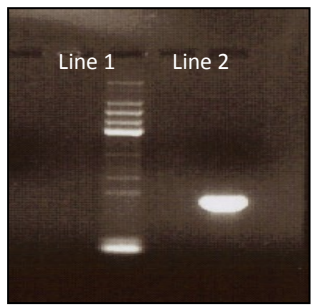

Figure 1: Image of PCR product running on 1.5 % of agarose gel for 4 hours at 50V. Line 1: Ladder, (NEB 100bp). Line 2: *invA* primer matched with *Salmonella* spp. DNA template at 284 bp

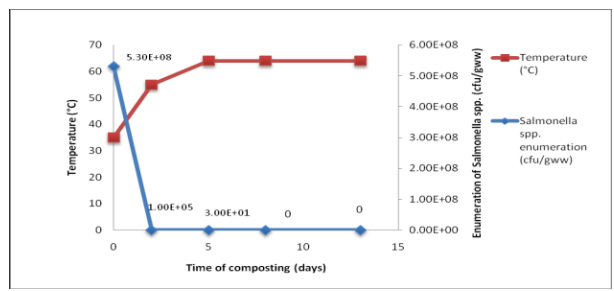

Figure 2: The Enumeration of *Salmonella* spp. and temperature profiles of the laboratory-scale of composting test

Overall, this study was contributed to the valuable knowledge about the inactivation of *Salmonella* spp. during composting by inoculums the solution that contained the *invA* gene sequences using molecular based characterization.

## 1. INTRODUCTION

*Salmonella* spp. is well known as pathogen indicator, in standard quality of composting. The role of *Salmonella* spp. in defining the ecological quality composts products was established in the European Commission Decisions. The UK composting standard (BSI, 2005) requires composts that are sold to be free of *Salmonella* spp. The standard cultivation was recommended by composting standard guideline as a technique to enumerate the *Salmonella* spp. (BS EN ISO 6579:2002) which may take as long as 2 days for completion. Nowadays, several of molecular tools were used in environmental study to provide more accurate and rapid results. PCR (polymerase chain reaction) makes possible for detect the numbering of cells in environmental samples, and not only those which are culturable. The behavior and detections of *Salmonella* spp. as pathogenic indicator using PCR was widely used in composted biosolid (Wéry et al., 2008, Novinscak et al., 2007) and waste water treatment (Wéry et al., 2008, Shannon et al., 2007, Kapley et al., 2001).

This study was designed in attempt to develop the molecular techniques to enumerate the *Salmonella* spp. during inactivation in composting. We reported here the selection on primers for PCR which was a crucial preliminary step in molecular techniques. The *invA* gene was utilized in this study as it was reported was the specific sensitivity in identification of a *Salmonella* strains is based on the amplification of 99.4% of the *invA* gene (De Clercq, 2007, Rahn et al, 1992).

In this work, the characteristic of *invA* gene sequences in *Salmonella* spp. solution was determined by using molecular techniques. The *Salmonella* spp. was then used as inocula solution for inactivation study of pathogen indicator in composting. This information's about *invA* gene would give a valuable knowledge for further developing molecular technique on enumeration of *Salmonella* spp. using *invA* gene in compost material.

## 2.0. MATERIAL AND METHODS

### 2.1 Bacterial cell culture and culture media

*Salmonella enteritidis* culti-loops, ATTC 13076 (Remel, Lenexa, KS) was suspended in tryptone soya broth (Oxoid). Tryptone soya broth (TSB) was prepared earlier by suspend 30g in 1 liter of distilled water into clean flask. The flask contained the broth was then sterile by autoclaved at 121°C for 15 min. The cell was grown in tryptone soya broth after gives enough time that is approximately 24 hours in fridge. The cell culture was determined by plating appropriate serial dilutions using sterile ringer solution on tryptone soya agar (TSA) from Oxoid. Tryptone soya agar was prepared by suspend 40g of tryptone soya agar in 1 litre purified water. Bring it to the boil until it dissolved completely and sterilized at 121°C for 15 min. The grown cells from tryptone soya broth (TSB) were counting by plating out on TSA immediately after the incubation for 24 hours at 37°C. The culture media that contained *Salmonella* spp. was then proceeds to *invA* gene characterisation using molecular techniques before used it as inocula solution in compost material preparation.

### 2.4 Selection of primers

The selection of primer for *Salmonella* spp. is summarized in Table 1. The specificity of the probes and primers at the species level were verified by performing the NCBI-BLAST2 (Basic Local Alignment Search Tool) by EMBL Nucleotide Sequences Database (European Molecular Biology Laboratory) from European Bioinformatics Institute (Wéry et al., 2008). The primer provided by Herrera-León, (2004) which use the Multiplex PCR-DNA distinguished the most common Phase-1 flagella antigens of *Salmonella* spp. This study needs more determination before either can be adapted into environmental samples material.

There have also been reports of primer being used in PCR; however, the study does not shown the specific targets of sequences of DNA (Chiu et al., 2005, Park, 2006, Suh and 2005) as they not mentioned which specific target of DNA in their study. Meanwhile, primers at sequence ST-11 for reverse and ST15 for forward (Aabo, 1993, Rijpens et al., 1999, Li, 2002) have a significant length for detection which is about 429 bp. The detection sensitivity of the PCR assay for the pure cultures was independent of the template preparation method (P=0.946) (Li, 2002).

However, the *invA* 139-R and *invA* 141-F primer of *Salmonella* spp. is more considerable for the PCR method. This primer was reported to successfully amplify 626 DNA of the 630 strains of *Salmonella* tested, when a single colony of bacteria was incorporated into a PCR mixture. Moreover, the specific sensitivity in identification of a *Salmonella* strains is based on the amplification of 99.4% of the *invA* gene (De Clercq, 2007, Rahn et al., 1992). This study proved much greater significant than ST-11 and ST-15 (Aabo, 1993, Rijpens et al., 1999, Li, 2002).

Table 1: **Summarized of review about selection of primer of *Salmonella* spp**

| Forward primer, 5'-3' | Reverse primer, 5'-3' | Author | NCBI-BLASTt2 nucleotide | Fragment length is produced |
|---|---|---|---|---|
| 547-569<br>GTG ATC TGA AAT CCA GCT TCA AG | 1055-1078<br>AAG TTT CGC ACT CTC GTT TTT GG | (Herrera-León, 2004) | >EM_PRO:Z15068; Z15068 S.enteritidis fliC gene for phase-1 flagellin (partial) Length = 1515.Score = 42.8 bits (46), Expect = 0.034.Identities = 23/23(100%) Strand = Plus / Plus Query: 1   gtgatctgaaatccagcttcaag 23 Sbjct: 548 gtgatctgaaatccagcttcaag 570 | 500 bp |
| *Salmonella* ST11<br>AGC CAA CCA TTG CTA AAT TGG CGC A | ST15<br>GGT AGA AAT TCC CAG CGG GTA CTG | (Rijpens et al., 1999), (Li, 2002), (Aabo, 1993) | >EM_PRO:AE008758; AE008758 *Salmonella typhimurium* LT2, section 62 of 220 of thecomplete genome.Length = 22470 Score = 46.4 bits (50), Expect = 0.003 Identities = 25/25(100%).Strand = Plus / Minus Query: 1 agccaaccattgctaaattggcgca 25 Sbjct: 116 agccaaccattgctaaattggcgca 92 Score = 46.4 bits (50), Expect = 0.003 Identities = 25/25(100%) Strand = Plus / Minus Query: 1 agccaaccattgctaaattggcgca 25 Sbjct: 2673 agccaaccattgctaaattggcgca 2649 | 429 bp |
| X68203, *phoE*<br>AGC GCC GCG GTA CGG GCG ATA AA | S50669<br>ATC ATC GTC ATT AAT GCC TAA CGT | (Kapley et al., 2001) | >EM_PRO:X68023; X68023 S.typhimurium phoE gene.Length = 1232.Score = 42.8 bits (46), Expect = 0.034.Identities = 23/23(100%).Strand = Plus / Plus Query: 1 agcgccgcggtacgggcgataaa 23 Sbjct: 831 agcgccgcggtacgggcgataaa 853 | 365 bp |
| Itsf<br>TAT AGC CCC ATC GTG TAG TCA GAA C | Itsr<br>TGC GGC TGG ATC ACC TCC TT | (Park, 2006), (Chiu et al., 2005) | No specific results in BLAST-NBCL2 | 312 bp |
| *Salmonella* spp. (ompC) S18<br>ACC GCT AAC GCT CGC CTG TAT | *Salmonella* spp. (ompC) S19<br>AGA GGT GGA CGG GTT GCT GCC GTT | (Burtscher, 2003) | >EM_PRO; AL627274 *Salmonella enterica* serovar Typhi *(Salmonella typhi)* strain CT18, complete chromosome; segment 10/20Length = 256050.Score = 39.2 bits (42), Expect=0.40.Identities = 21/21(100%)Strand | 159 bp |
| SAF<br>TTG GTG TTT ATG GGG TCG TT | SAR<br>GGG CAT ACC ATC CAG AGA AA | (Suh and 2005) | >EM_PRO:M90846; M90846 *Salmonella typhimurium* InvA *(invA)* gene, complete cds.Length = 2176.Score = 37.4 bits (40), Expect | 298 bp |

| | | | = 1.5.Identities = 20/20(100%) Strand = Plus / Plus Query: 1   ttggtgtttatggggtcgtt 20 Sbjct: 232 ttggtgtttatggggtcgtt 251 | |
|---|---|---|---|---|
| *invA* 139. M90846 GTG AAA TTA TCG CCA CGT TCG GGC AA | *invA* 141 TCA TCG CAC CGT CAA AGG AAC C | (De Clercq, 2007);(Rahn et al., 1992), (Gunaydin, 2007), (Malorny et al., 2003) | >EM_PRO: U43250 *Salmonella enterica* invasion protein (*invA*) gene, partialcds. Length = 1950.Score = 44.6 bits (48), Expect = 0.009.Identities = 24/24 (100%) Strand = Plus / Plus | 284 bp |
| *invA* gene 167-186 TCG TCA TTC CAT TAC CTA CC | *invA* gene 234-285 AAA CGT TGA AAA ACT GAG GA | (Hoorfar, 2000); (Wéry et al., 2008) | >EM_PRO:M90846; M90846 *Salmonella typhimurium* InvA (*invA*) gene, complete cds.Length = 2176.Score = 37.4 bits (40), Expect = 1.5.Identities = 20/20(100%) Strand = Plus / Plus Query: 1   tcgtcattccattacctacc 20 Sbjct: 167 tcgtcattccattacctacc 186 | 119 bp |
| *invA* gene GCG TTC TGA ACC TTT GGT AAT AA | CGT TCG GGC AAT TCG TTA | (Daum, 2002); (Novinscak et al., 2007) | >EM_PAT; BD340480 Primers for *Salmonella* spp.  Detection and Detection of *Salmonella* spp. Using the Primers. Length = 497.Score = 41.0 bits (44), Expect = 0.11.Identities = 22/22 (100%)Strand = Plus / Minus | 102 bp |
| U43272 CGT TTC CTG CGG TAC TGT TAA TT | AGA CGG CTG GTA CTG ATC GAT AA | (Shannon et al., 2007); (Lee, 2006) | >EM_PAT; BD340480 Primers for *Salmonella* spp.  Detection and Detection of *Salmonella* spp. Using the Primers. Length = 497 Score = 42.8 bits (46), Expect = 0.033Identities = 23/23 (100%) Strand = Plus / Plus | 67 bp |

After being through several reviews about primer of *Salmonella* spp. in molecular techniques (shown in table above) the selection of the best primer was made. PCR primers were selected that target a 285 bp segment of *invA* gene of *Salmonella* (the *invA* gene of *Salmonella* is a component of the cell invasion apparatus). These primers shown in Table 2 were *invA*139 and *invA*141 which target locations 287-312 and 571-550 within the *invA* gene, respectively (Rahn et al., 1992). According to Rahn et al.,(1992) these primers have been shown to have excellent specificity for *Salmonella*, detecting 99.4% *Salmonella* strains without false positives when tested against 630 *Salmonella* strains and 142 non-*Salmonella* strains.

Table 2: **Base sequences and locations of oligonucleotide primers (Rahn et al., 1992)**

| Primer | Oligonucleotide sequence (5'-3') | Location within *invA* gene |
|---|---|---|
| **139** | gtgaaattatcgccacgttcgggcaa | 287-312 |
| **141** | tcatcgcaccgtcaaaggaacc | 571-550 |

A PCR with these primers was conducted with DNA from *Salmonella enteritidis* (ATCC13076). A suspension of *Salmonella* cells in sterile distilled water were lysed by heating to 99°C for 5 minutes. Cell debris was removed by centrifugation. A PCR reaction was set-up containing 2.5 µl of DNA solution, 5 units of GoTaq DNA polymerase (Promega Corp., USA), 1× GoTaq PCR reaction buffer (containing 1.5mM $MgCl_2$), 0.2mM PCR nucleotide mix (Promega Corp., USA), and 0.6 µM DNA primers in a final volume of 50 µl. This reaction mixture (and a sterile control) was incubated at 95°C for 2 min, and then cycled 30 times through three steps: denaturing (95°C, 30s), annealing (50°C, 30s), primer extension (72°C, 45s). This was followed by a final extension step at 72°C for 7min. The PCR product was purified using agarose gel electrophoresis and a QIAquick Gel Extraction Kit (QIAGEN Ltd, UK). The PCR product, which was just under 300bp long, was sent for direct DNA sequencing (ABI 3100*xl* Capilliary Sequencer) using both the *invA*139 and *invA*141 primers.

## 2.5 Purification of Salmonella spp. DNA

The result from PCR for *Salmonella* spp. was then proceeds for purification and extraction of DNA for sequencing of *invA* gene characterisation. The PCR produced the expect fragment length which is 284 bp for *Salmonella spp.* (D. De Clercq, 2007) refer to results and discussion section (shown in Figure 1). The fragment produced by the *Salmonella* spp. primers was isolated on a agarose TBE gel fragment was isolated with a purification kit due to clarity of band that had been shown.

The PCR purification spin protocol for *Salmonella* spp. fragment was utilised at laboratory by using QIAquick PCR Purification Kit protocol. This protocol was designed to purify single-or double- stranded DNA fragments from PCR and other enzymatic reaction. 5 volumes of buffer PBI was added to 1 volume of *Salmonella* spp. PCR sample and mix. The colour of the mixture should be yellow which is similar to Buffer PBI without the PCR sample. A 2ml collection tube was used in to place in a QIAquick spin column. Then the sample was applied to the QIAquick column and centrifuged for 30-60s, to bind DNA. QIAquick column was washed by added 0.75ml of Buffer PE and centrifuged again for 30-60s. The Buffer PE was added with 96-100% ethanol before to be used. The residual ethanol from Buffer PE was completely removed by additional of centrifugation about 1 min repeated twice. Clean and sterile of 1.5 ml centrifuge tube was used to place QIAquick column. 50µ of Buffer EB (10mM Tris-Cl, pH 8.5) was added to the centre of the QIAquick membrane and then centrifuged for 1 min to elute the DNA. This centrifuge tube was stored in -20°C for next further sequencing protocol.

## 2.6 Sequencing of DNA

The concentration of DNA fragment for *Salmonella* spp. was measured by using NanoDrop ND-1000 UV-spectrophotometer (Thermo-Fisher Scientific Inc., USA). This process was required before the sequencing protocol. The result for this measurement is 21.82 ng/l of *Salmonella* spp. The fragment DNA was diluted to required concentration for sequencing in a ABI 3100 Capillary Sequencing machine. The sample was prepared about 10-12 µl of template at a concentration of 1.00 ng/µl (200-500 bp). Both primers (forward and reverse) of *Salmonella* spp. were prepared about 4-6 µl at a concentration of 1.6 pmol/µl. All of the diluted templates and primers were placed in sterile small ependorf. Table 4 described in results and discussion section shown the results of DNA sequences from ABI 3100 Capillary Sequencing machine.

## 2.7 DNA sequence analysis

Sequences analysis was using the NCBI-BLAST2 program and the EMBL release nucleotide database (Wéry et al., 2008). Default settings were used for the BLAST parameters (match/mismatch scores 2, -3, open gap penalty 5, gap extension penalty 2). The NCBI-BLAST2 program also was running for several types of primers during selection of *Salmonella* spp. primer. The results was indicates the specificity of *Salmonella* spp. in BankGene library (shown in Table 3 in results and discussion section).

## 2.8 Polymerase chain reaction procedure

Several PCR experiments were carried out on pure cell cultures to validate the functionality of the reagents and primers. The cells of *Salmonella* spp. were taken from an individual colony on the surface of an agar plate using a sterile toothpick and resuspended in 100µl of sterile distilled water. Samples were heated at 99 °C for 5 minutes and centrifuged at a top speed in a micro centrifuge for 1 minute to remove cell debris. The supernatant was then transferred into a new tube to be used as a source of DNA.

The PCR reaction mixture contained 2.5µl of DNA solution from the procedure above, 5 units of GoTaq reaction buffer (from Promega Corp., USA), 1 x PCR reaction buffer, 1.5mM $MgCl_2$ (already in the GoTag reaction buffer), 10 mM PCR nucleotide mix (Promega Corp., USA), and 1.5 µM DNA primer in a final volume of 50 µl. The reaction mixtures were incubated at 95°C for 2 min, and then cycled 30 times for another three steps: denaturing (95°C, 30s), annealing (50°C, 30s), primer extension (72°C, 45s). This was followed by final extension step at 72°C for 7 min. Amplification product sizes were verified by electrophoresis of 10 µl samples in a 1.0% agarose TBE gel with ethidium bromide straining (Stewart et al., 2007).

### 2.9 Raw material for compost material

The preparation of compost material was made after the molecular characterisation of *invA* gene in *Salmonella* spp. solution. Compost material was prepared by adding inocula solutions contained *Salmonella* spp. with matured compost produced from kitchen waste. This composting mixture was adjusted to appropriate ratio at 6:4 (w/w). The compost mixture was shredded to an average size of 5 – 20 mm (Hu et al., 2009) before it was placed in a laboratory compost reactor.

### 2.10 Laboratory-scale composting

The laboratory composting apparatus used was the same as that used in the DR4 biodegradability tests method devised by WRc (Waste Research Centre) in the United Kingdom. The material (compost mixture) was incubated under aerobic conditions in a reactor using forced aeration. The composting reactor was cylindrical (22 cm height x 8 cm diameter) with a perforated plate at the bottom to distribute the air supplied, and was loaded with 400 g of the mixture. The air was supplied using a pump at a constant flow rate of 0.5L/min, which was measured and controlled using a flow meter (tube-and-float type). In this sealed reactor, the air was forced up through the compost material, and passed out of the reactor. Then the air was passed through a condenser, to remove surplus liquid before exhausting to atmosphere. The temperature of the compost material was monitored with a thermocouple which was inserted in the sample.

### 2.11 Conventional enumeration approach

The conventional method for the enumeration of *Salmonella* spp. in composting was carried out using serial dilution followed by a standard membrane filtration technique. This method is recommended by the UK compost quality standard method PAS 100 (BSI, 2005). The enumeration of *Salmonella* spp. (BS EN ISO 6579:2002) used the Muller-Kauffmann tetrathionate/novobiocin broth as a resuscitation medium followed by growth on Rambach agar. Compost sample preparation involved taking a 25 g sub-sample and placing this into a stomacher bag together with 225ml of sterile PBS (phosphate buffered saline). This sample was then stomached for 60 seconds and subject to serial 10-fold dilution of the resulting supernatant using PBS.

## 3. RESULTS AND DISCUSSION

### 3.1 16S rRNA Gene sequencing: selecting primer

The desired PCR product was approximately about 284 bp for *Salmonella* spp. in length. This sets of PCR primer were selected from previous studies (McDaniels, 1996) was described in Table 3.

Specificity of the all probes and primers at the species level were verified by performing the NCBI-BLAST2 (Basic Local Alignment Search Tool) by EMBL Nucleotide Sequences Database (European Molecular Biology Laboratory) from European Bioinformatics Institute. Default setting were used for the BLAST parameters (match/mismatch scores 2,-3, open gap penalty 5, gap extension penalty 2). The sensitivity of identification of *Salmonella* strains is based on the amplification of the *invA* gene in this primer was 99.4 %.

Table 3: **DNA sequences used for PCR primer and probes**

| *Primer and probes* | *Sequences (5'→3')* | *Microorganisms* | *Genes* | *Functions* | *GenBank accession no.* |
|---|---|---|---|---|---|
| Sal-F | GTG AAA TTA TCG CCA CGT TCG GGC AA | *Salmonella* spp. | *invA* | Inavasion protein | M90846 |
| Sal-R | TCA TCG CAC CGT CAA AGG AAC C | *Salmonella* spp. | *invA* | Inavasion protein | M90846 |

### *3.2 Specificity of invA gene in Salmonella spp.*

The primers gave a good result and matched correctly with *Salmonella* spp. DNA template. The source of DNA from broth solution was prepared before the experiment. Figure 1 showed the PCR product outcome after within 4 hours running on 1.5% of agarose gels. The gel was needed to run out until the bottom of the gel (required longer time) so the band appeared until the end of product. The result showed that *Salmonella* spp. (DNA template) matched and bond with the *invA* gene primer (Line 2) at 284 bp. This value was same as stated by D. De Clercq, (2007). This result was shown a very clear band indicates high specificity and matched of *invA* gene in *Salmonella* spp. solutions. Thus, these primers were suitable for next further steps such as purification and sequencing for characterisation of *invA* gene in *Salmonella* solutions. The *Salmonella* spp. solution was proved contained the *invA* gene then will be used in composting for pathogen indicator inactivation study in composting trial.

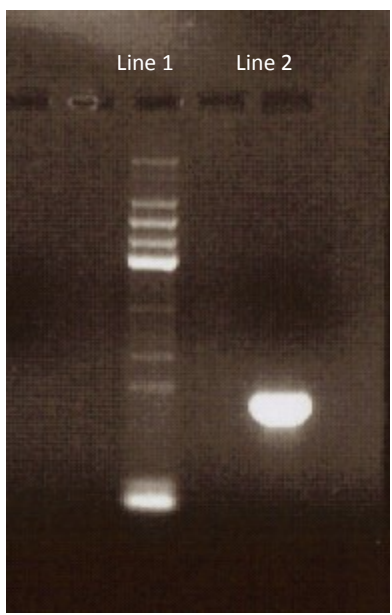

Figure 1: **Image of PCR product running on 1.5 % of agarose gel for 4 hours at 50V. Line 1: Ladder, (NEB 100bp). Line 2: *invA* primer matched with *Salmonella* spp. DNA template at 284 bp.**

### *3.3 Sequencing of Salmonella spp. DNA*

The concentration of DNA fragment for *Salmonella spp.* was measured by using NanoDrop ND-1000 UV-spectrophotometer (Thermo-Fisher Scientific Inc., USA). The result for this measurement is 21.82 ng/l of *Salmonella spp*. The fragment DNA was diluted to required concentration for sequencing in ABI 3100 Capillary Sequencing machine. Table 4 below are the results of DNA sequences from ABI 3100 Capillary Sequencing machine. These results provided valuable knowledge of characterisation using molecular techniques using *invA* gene primer for *Salmonella* spp.

Table 4: **Sequencing results for forward reverse primers of *Salmonella spp.***

| *Primer* | *Result of DNA sequences* |
|---|---|
| *Salmonella* spp. forward | >D12_008_*Salmonella*-sp_Sal-F.ab1 TNNNNTTTTTNNNNTGGTCTTCTCTATTNNCACCGTGGTCCAGTTTATCGTTATTAC CAAAGGTTCAGAACGCGTCGCGGAAGTCGCGGCCCGATTTTCTCTGGATGGTATGC CCGGTAAACAGATGAGTATTGATGCCGATTTGAAGGCCGGTATTATTGATGCGGAT GCTGCGCGCGAACGGCGAAGCGTACTGGAAAGGGAAAGCCAGCTTTACGGTTCCT TTGACGGTGCGATGAA |
| *Salmonella* spp. reverse | >A03_001_*Salmonella*-sp_Sal-R.ab1 NNNNNNNNNNGCCGTTCGCGCGCAGCATCCGCATCAATAATACCGGCCTTCAAATC GGCATCAATACTCATCTGTTTACGGGCATACCATCCAGAGAAAATCGGGCCGCGAC TTCCGCGACGCGTTCTGAACCTTTGGTAATAACGATAAACTGGACCACGGTGACAA TAGAGAAGACAACAAAACCCACCGCCAGGCTATCGCCAATAACGAATTGCCCGAA CGTGGCGATAATTTNNN |

## 3.4 Inactivation of Salmonella spp. in composting

The *Salmonella* spp. matched the specificity with *invA* gene (accordingly to molecular characterization results) was then used as inocula solution in composting trial. This would provide the results about inactivation of *Salmonella* spp. in laboratory scale of composting. This change in the numbers of *Salmonella* spp. was enumerated using the conventional method (membrane filtration) was shown in Figure 2. The Figure 2 also shows the temperature profiles of the laboratory-scale composting test. The temperature was increased from ambient to the thermophilic range over a 2 day period. It was then kept at that temperature until day 13. According to Deportes et al., (1995) temperatures of approximately 55-60 °C for at least 3 days are recommended to be effective against *Salmonella* spp. The temperature profiles in the laboratory scale composters were set to at least exceed this level of exposure as can be seen from Figure 2. The laboratory scale study showed that after composting for 8 days the numbers of *Salmonella* spp. were below the limits in the UK compost standard (PAS 100) which requires the compost to be free of *Salmonella* spp.

This study is the preliminary trial on *invA* gene for *Salmonella* spp. inactivation study of composting trial. The results has been showed that the *invA* gene was easily recognize in *Salmonella* spp. by using PCR techniques.The knowledge from characteristic of *invA* gene provided in this study will further used as a tools in molecular techniques for enumeration of pathogen indicator in composting. According to Rahn et al. (1992) the *invA* gene contains sequences unique to and that primers produced to target this region can be used to differentiate between *Salmonella* and other organisms. Meanwhile in this study, the reduction of *Salmonella* spp. that confirmed contained a sequences of *invA* gene was significantly used as inocula solution in composting. The reduction of *Salmonella* spp. in composting trial was meets the standard compost guideline however the enumeration provided was not specific to *invA* gene. This is because the technique used in this study was a standard recommended technique that is membrane filtration. Therefore, the molecular techniques on enumeration of *Salmonella* spp. contained with *invA* gene should be further investigated. The conventional techniques in this study however was necessarily to confirm the inactivation *Salmonella* spp. prepared as inocula solution mixed with compost material. In some case, the enumeration of *Salmonella* spp. using molecular techniques would involve pro and contras. Reported by Wéry et al.(2008) the environment sample always associated with the inhibitory effect and contaminants compounds that present in samples. Thus the conventional method described in this study is important to collaborate with further developing molecular techniques on enumeration of *Salmonella* spp. that associated with sequences of *invA* gene in compost material.

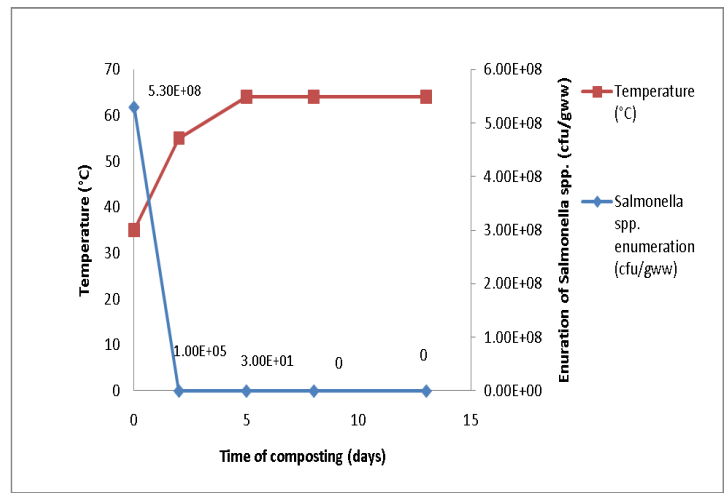

Figure 2: **The Enumeration of *Salmonella* spp. and temperature profiles of the laboratory-scale of composting test**

## 4. CONCLUSION AND RECOMMENDATIONS

This study was successfully determined the suitable primers set with *Salmonella* spp. DNA. The *invA* gene was selected as forward and reverse primers due to it's provided a highly specific matched with *Salmonella* spp. The PCR amplification for *invA* primers and *Salmonella* spp DNA was produced significantly good results. Thus, it was suitable to be used for further investigation for inactivation of pathogen indicator using molecular techniques along with standard conventional procedures.

## 5. ACKNOWLEDGMENT

This study was funded by Ministry of Higher Education of Malaysia and University of Tun Hussien Onn Malaysia. The authors wish to thank the Public Health Laboratory staff, School of Civil Engineering, University of Leeds for their valuable support and excellent laboratory facilities.